\documentclass[usegraphicx]{mn2e}

\title{Evidence of fast rotation in dwarf elliptical galaxies}
\author[S.~Pedraz et al.]
{S.~Pedraz,$^{1,2}$\thanks{E-mail: spm@astrax.fis.ucm.es}
J.~Gorgas,$^{1}$
N.~Cardiel,$^{1}$
P.~S\'{a}nchez-Bl\'{a}zquez$^{1}$
and R.~Guzm\'{a}n$^3$\\
$^{1}$Departamento de Astrof\'{\i}sica, Facultad de Ciencias
F\'{\i}sicas, Universidad Complutense de Madrid, E28040--Madrid, Spain\\
$^{2}$Calar Alto Observatory, CAHA, Apdo. 511, 04004 Almer\'{\i}a, Spain\\
$^{3}$Department of Astronomy, University of Florida, P.O. Box 112055, 
Gainesville, FL 32611-2055}
\date{Accepted 2002 April 9.
      Received 2002 April 1;
      in original form 2001 December 20}

\pagerange{\pageref{firstpage}--\pageref{lastpage}}
\pubyear{}

\newcommand{\reduceme}{\mbox{R\raisebox{-0.35ex}{E}D%
\hspace{-0.05em}\raisebox{0.85ex}{uc}\hspace{-0.90em}%
\raisebox{-.35ex}{{m}}\hspace{0.05em}E}}

\begin{document}

\maketitle

\label{firstpage}

\begin{abstract}
In this Paper we investigate the kinematical properties of early-type dwarfs
by significantly enlarging the scarce observational sample so far available.
We present rotation curves and mean velocity dispersions for four bright
dwarf ellipticals and two dwarf lenticular galaxies in the Virgo cluster.
Most of these galaxies exhibit
conspicuous rotation curves. In particular, five out of the six new galaxies
are found to be close to the predictions for oblate spheroids flattened by
rotation. Therefore, and contrary to the previous observational hints, the
present data suggest that an important fraction of dwarf early-type galaxies
may be rotationally supported.

\end{abstract}

\begin{keywords}
galaxies: dwarf -- galaxies: elliptical and lenticular, cD -- galaxies:
kinematics and dynamics
\end{keywords}

%%%%%%%%%%%%%%%%%%%%%%%%%%%%%%%%%%%%%%%%%%%%%%%%%%%%%%%%%%%%%%%%%%%%%%%%%%%%%%%
\section{Introduction}

Although dwarf elliptical galaxies 
(dEs) constitute the dominant galaxy population in nearby clusters, their
origin and nature are still unclear. One of the main questions under debate is
whether dEs represent the extension of the ``classical'' ellipticals (Es) to
lower luminosities, or, on the contrary, whether they are the result of
distinct formation and evolution processes (for a review see, e.g., Ferguson
\& Binggeli 1994). While early observational works suggested a dichotomy in
the structural properties of both galaxy families (e.g. Kormendy 1985),
Binggeli \& Jerjen (1998) (see also Jerjen \& Binggeli 1997) have shown that,
if a S\'{e}rsic law is used to fit the observed luminosity profiles, dEs
appear as the true low-luminosity extension of the classical Es. Previous
workers have also studied the possible dichotomy between dEs and Es by
comparing the flattening distributions (e.g. Ryden \& Terndrup 1994),
clustering properties (Conselice, Gallagher \& Wyse 2001) and stellar
populations (Gorgas et al. 1997, hereafter G97; Drinkwater et al. 2001) of dEs
in nearby clusters. Another key issue that should provide important clues to
the open question of the possible dichotomy is the analysis of rotational
properties.

%\footnote{Note that in G97 we
%employed the nomenclature of Kormendy \& Bender~(1994), where dwarf
%ellipsoidal galaxies such as NGC~205 were called ``spheroidal'' instead of
%``dwarf elliptical''.  However, here we have preferred the nomenclature of
%``dwarf elliptical'', as in Binggeli~(1994).}

From the work of Davies et al. (1983) (see also Halliday et al. 2001), it is
well known that low-luminosity E galaxies, in contrast with giant Es, are
supported by rotation. If dEs were the extension to even lower luminosities of
the classical Es, one should expect to find relatively high rotational
velocities along the major axis of these objects. The observational data in
this subject require high signal-to-noise (S/N) spectra at faint levels of
surface brightnesses and it is, therefore, very scarce. Until recently,
rotation curves had only been measured for 6 dEs: two dwarfs in Virgo (VCC~351
and IC~794; Bender \& Nieto 1990, hereafter BN90), the 3 dwarfs companions of
M31 (e.g. Bender, Paquet \& Nieto 1991) and the Fornax dE (Mateo et
al. 1991). In all these cases, dEs were found to rotate too slowly to be
consistent with an oblate, isotropic body flattened by rotation. Very
recently, new data have been added to this sparse statistics. On the one hand,
Geha, Guhathakurta \& van der Marel (2001, hereafter GGV01) have presented
kinematical data for a sample of 4 Virgo dEs, finding that, in agreement with
the previous results, they are slow rotators. On the other hand, De Ricke et
al. (2001, hereafter DR01) have obtained deep spectroscopic data of the dE
FS76 (in the NGC 5044 group), showing that the galaxy is rotating as fast as
predicted by the isotropic models. In this Paper we revisit the question of
rotation in dEs by practically doubling the observational sample so far
available. We will show that most of the galaxies in the new sample tend to be
rotationally supported and, therefore, fast rotation in dwarf elliptical
galaxies is not a rare event, as previously thought.

\begin{scriptsize}
\begin{table*}
\centering{
\caption{Kinematic data for the galaxy sample. Galaxy types have been taken
from Binggeli, Sandage \& Tammann (1985). Absolute magnitudes have been
computed using a distance modulus to Virgo of $-30.82$, which corresponds to
$H_0=75$~km~s$^{-1}$~Mpc$^{-1}$. PA is the slit position angle (north to
east), which corresponds to the major axis position according to the
following references: (1) Jedrzejewski (1987); (2) Paturel et al. (1989), and
(3) Binggeli \& Cameron (1993).
Sources for ellipticities ($\varepsilon$)
are: (3) Binggeli \& Cameron (1993); (4) Mean from Binggeli et al. (1985) and Djorgovski (1985);
(5) Ryden et al. (1999), and (6) GGV01. 
The central velocity dispersions ($\sigma_0$ , in km~s$^{-1}$) correspond to a
$2^{\prime\prime}\times4^{\prime\prime}$ central aperture. Mean velocity
dispersions ($\overline\sigma$), maximum rotational velocities ($V_{\rm max}$)
and the anisotropy parameters ($(V/\sigma)^\ast$), together with their formal
errors, have been derived as explained in the text.}
\label{table_sample}
\begin{tabular}{llrr@{\ }cr@{\ }cr@{$\ \pm\ $}rr@{$\ \pm\ $}rr@{$\ \pm\ $}rr@{$\ \pm\ $}r}
\hline
\multicolumn{1}{c}{Galaxy} & \multicolumn{1}{c}{Type} & 
\multicolumn{1}{c}{$M_{\rm B}$} & 
\multicolumn{1}{c}{PA} & Source &
\multicolumn{1}{c}{$\varepsilon$} & Source &
\multicolumn{2}{c}{$\sigma_0$} & 
\multicolumn{2}{c}{$\overline\sigma$} & 
\multicolumn{2}{c}{$V_{\rm max}$} & 
\multicolumn{2}{c}{$(V/\sigma)^\ast$}\\ \hline
NGC 4415 & d:E1,N   & $-17.10$ &   0 & 2 & 0.11 & 3 & 42.7 &  4.2 & 41.7 &  2.2 & 20.9 &  1.2 & 1.43 & 0.38 \\
NGC 4431 & dS0(5),N & $-17.12$ & 177 & 2 & 0.42 & 5 & 41.9 &  4.9 & 52.9 &  2.1 & 32.0 &  6.3 & 0.71 & 0.16 \\
NGC 4489 & S0$_1$(1)& $-17.98$ & 160 & 1 & 0.08 & 4 & 51.1 &  1.8 & 50.3 &  1.4 & 22.8 &  8.1 & 1.55 & 0.73 \\
IC 794   & dE3,N    & $-16.62$ & 110 & 2 & 0.24 & 6 & 41.1 &  3.2 & 43.4 &  2.5 &  3.4 &  1.7 & 0.14 & 0.07 \\
IC 3393  & dE7,N    & $-16.07$ & 133 & 3 & 0.52 & 5 & 25.0 &  7.6 & 26.8 &  4.3 & 18.0 &  1.7 & 0.65 & 0.14 \\
UGC 7436 & dE5      & $-16.02$ & 120 & 2 & 0.44 & 5 & 35.0 &  4.1 & 36.7 &  2.8 & 14.9 &  3.1 & 0.47 & 0.11 \\
\hline
\end{tabular}
}
\end{table*}
\end{scriptsize}

%\footnote{Note that in Gorgas et al.~(1997) we employed the
%nomenclature of Kormendy \& Bender~(1994), where dwarf ellipsoidal galaxies
%such as NGC~205 were called ``spheroidal'' instead of ``dwarf elliptical''.
%However, here we have preferred the nomenclature of ``dwarf elliptical'', as
%in Binggeli~(1994).}
%%%%%%%%%%%%%%%%%%%%%%%%%%%%%%%%%%%%%%%%%%%%%%%%%%%%%%%%%%%%%%%%%%%%%%%%%%%%%%%
\section{Galaxy sample, observations and data reduction}

The galaxies analysed in this work correspond to those already presented in
G97. This sample of Virgo galaxies, which is listed in
Table~\ref{table_sample}, comprises 4 dwarf ellipticals and two dwarf
lenticular galaxies.  Although NGC~4489 is classified as S0 (non dwarf) in
Binggeli, Sandage \& Tammann (1985), it exhibits a mean surface brightness
within the effective radius of $SB_e=22.24$~mag~arcsec$^{-2}$ (Faber et
al. 1989), which, in the $SB_e$--absolute magnitude plane, places it closer to
the locus of dwarf ellipticals than to the low-luminosity early-type
galaxies\footnote{At the absolute magnitude of NGC~4489, classical ellipticals
and S0's attain typical values of $SB_e$ between 18 and 21~mag~arcsec$^{-2}$,
whereas for the dwarf galaxies of this sample $22.5<SB_e<23$
mag~arcsec$^{-2}$.}. It must be noted also that the present sample is biased
toward the brightest dwarf early-type galaxies of the Virgo cluster. 

Long-slit spectroscopic observations along the major axes of the galaxies were
carried out with the INT at La Palma.  Although the relevant observational
parameters were already described in G97, we want to highlight that
high-quality spectra, covering the 4700--6100~\AA\ range with a spectral
resolution of $\sigma=60$~km s$^{-1}$ (${\rm FWHM} \sim 2.5$~\AA) were
obtained out to, typically, the effective radii of the galaxies.

The data reduction and analysis were performed using the \reduceme\ package
(Cardiel et al.~1998; Cardiel 1999), specially developed for the parallel
treatment of data and errors in the reduction of spectroscopic
observations. In this sense, each fully processed spectrum is accompanied by
an associated error spectrum, which contains the propagation of the initial
random errors (due to photon statistics and read-out noise) throughout the
cascade of arithmetic manipulations in the reduction procedure. Full details
concerning the data reduction will be presented in Pedraz~(2002).

We complemented the observations of the targets with a large
set (39) of template stars from the Lick/IDS stellar library (Gorgas et
al.~1993; Worthey et al.~1994). The template list was selected to provide an
appropriate coverage in stellar atmospheric parameters ($T_{\rm eff}$ from
3247 to 6575~K, $\log g$ from 1.11 to 4.84, and [Fe/H] from $-2.15$ to 0.35),
suitable for the modelling of old stellar populations.

\section{Measurement of kinematic data}

\begin{figure}
\centering
\includegraphics[angle=-90,width=\columnwidth]{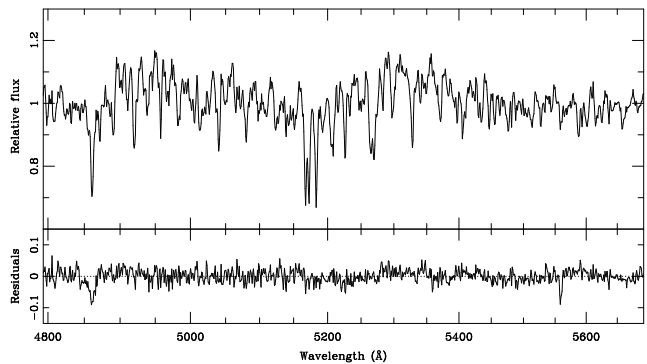}
\caption{Upper panel: Central spectrum of NGC~4415. 
The spectrum has been continuum
subtracted and normalized. Bottom panel: Residuals of the galaxy-composite
template fit, corresponding to the computed kinematic parameters ($\gamma=0.814$, $V_0=926.5$
km~s$^{-1}$ and $\sigma=42.7$ km~s$^{-1}$). The scale is the same than in the upper plot.}
\label{temp}
\end{figure}

\begin{figure*}
\centering
\includegraphics[angle=-90,width=\textwidth]{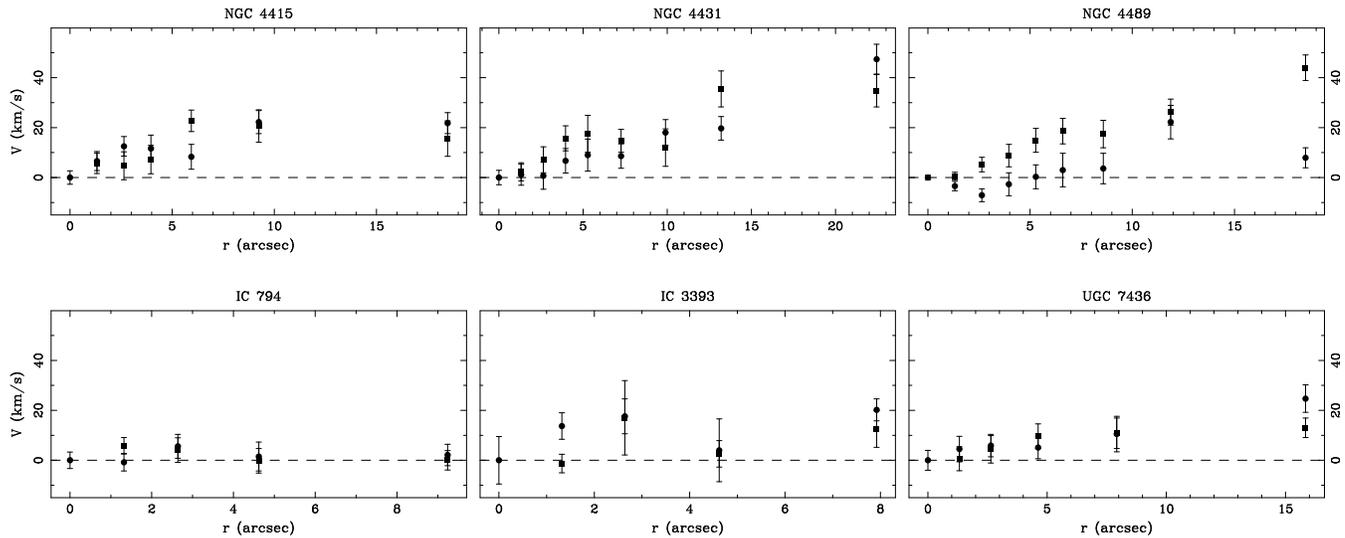}
\caption{Rotational velocities $V$ against
radius $r$ in arc seconds for the galaxy sample ($1^{\prime\prime}$
corresponds to $\sim 71$~pc).  The profiles are folded around the center, with
circles and squares corresponding to different galaxy sides.}
\label{panel}
\end{figure*}

In order to estimate radial velocities and velocity dispersions for each
galaxy spectrum, we wrote a dedicated program within \reduceme\ which follows
the MOVEL and OPTEMA algorithms described by Gonz\'{a}lez (1993). The MOVEL
algorithm, an improvement of the classic Fourier quotient method by Sargent et
al. (1977), is an iterative procedure in which a galaxy model is processed in
parallel to the galaxy spectrum. In this way, a comparison between the input
and recovered broadening functions for the model allows to correct the galaxy
power spectrum from any imperfections of the data handling in Fourier space.
The main improvement of the procedure is introduced through the OPTEMA
algorithm, which is able to overcome the typical template mismatch
problem. First, the 39 individual stellar spectra of the observing run were
binned to create a set of 8 template spectra corresponding to different
spectral types and luminosity classes (namely F0V, G0V, K0V, M0V, G5III,
G8III, K0III, K5III). For each galaxy central spectrum, these 8 template
spectra are scaled, shifted and broadened according to a first guess of the
$\gamma$ (mean line strength), $V$ (radial velocity)and $\sigma$ (velocity
dispersion) parameters. The next step is to find the linear
combination of the template spectra that best matches the observed galaxy
spectrum. This provides a first composite template which is fed to the
MOVEL algorithm. The output kinematic parameters are then used to create an
improved composite template and the process is iterated until it
converges. This iterative approach then provides an optimal template while
simultaneously computing the radial velocity and velocity dispersion of the
galaxy spectrum. In Figure~\ref{temp} we show a typical fit between the
central spectrum of a galaxy of the sample and its corresponding optimal
template corrected with the derived kinematic parameters. It must be noted
that, once an optimal template is derived for the central spectrum of a given
galaxy, this same template is used to derive the kinematics of the different
spectra along the galaxy radius.  This is a fair approximation since these
galaxies exhibit relatively flat stellar population gradients along the radius
(G97).

For each galaxy spectrum, random errors in the derived kinematic parameters
were computed by numerical simulations. In each simulation, a bootstrapped
galaxy spectrum, computed from the error spectra provided by the reduction
with \reduceme\ by assuming Gaussian errors, is fed to the algorithms
described above (note that a different optimal template is computed in each
simulation). Errors in the parameters are then calculated as the unbiased
standard deviations of the different solutions. These final errors are
expected to be quite realistic, since they incorporate all the uncertainties
of the whole reduction process, from the first reduction steps
(e.g. flatfielding) to the final measurements of the kinematics.

In order to assess the sensitivity of the above fitting procedure to measure
velocity dispersions of only a fraction of the spectral resolution
($\sigma_{\rm instr}$), we carried out several Monte Carlo simulations in
which we added Gaussian noise to artificially broadened stars. To summarize,
we conclude that, in order to measure a velocity dispersion of
$\sigma_{\rm instr}/2$ (30 km/s for our instrumental setup) with a relative
error below 50 \% ($< 15$ km/s) we needed galaxy spectra with a minimum S/N
(per \AA) of 20--25.

Central velocity dispersions $\sigma_0$, together with their corresponding
random errors, for the galaxy sample are listed in Table~\ref{table_sample}.
We must note that S/N for the central spectra ranged from 38 to 98. The
central velocity dispersion of NGC~4489 has been already measured by several
authors. Our result ($\sigma_0=51.1\pm1.8$) agrees with some of the previous
measurements, like those of Davies et al. (1983; $\sigma_0=53\pm15$), Faber et
al. (1989; $\sigma_0=49\pm7$) and Gonz\'{a}lez (1993; $\sigma_0=48\pm3$), but
it is somewhat smaller than the values given by Simien
\& Prugniel (1997; $\sigma_0=63\pm13$) and, especially, Smith et al. (2000;
$\sigma_0=72\pm7$). We must note that the spectral resolution of these two
last works ($\sigma_{\rm inst}\approx100$ km~s$^{-1}$) is dangerously high to
derive accurate velocity dispersions at these levels. IC~794 has also been
included in previous kinematics works. Our central velocity dispersion
($\sigma_0=41.1\pm3.2$) is marginally smaller than the value obtained by BN90
($\sigma_0=52\pm6$) but it perfectly agrees with the recent result of GGV01
($\sigma_0=41.6\pm0.9)$\footnote{This value has been computed, after the
digitization of Fig.~1 in GGV01, as the error weighted mean of all the
measurements with $r<2^{\prime\prime}$.}. Central velocity dispersions for
NGC~4431, IC~3393 and UGC~7436 have already been listed in the compilation by
Bender et al. (1992). We must note that these values (68, 55 and 45 km/s
respectively) are significantly larger than our measurements (42, 25 and
35~km/s). The original source of those values is unpublished data from
R.~Bender and, therefore, we do not have information enough to untangle the
origin of this discrepancy. In any case, given the high S/N ratios of our
spectra and the comparison with the previous published works, we are confident
that our velocity dispersions are free of significant biases.

Folded rotation velocity curves along the
major axes of the galaxies are presented in Figure~\ref{panel}. For the outer
regions, we co-added a sufficient number of spectra in the spatial direction
to guarantee a minimum signal-to-noise per \AA\ of 20. In the case of
NGC~4415, NGC~4431 and UGC~7436 the profiles extend out to the effective
radius of the galaxies, whereas, for the other three galaxies, the data reach
roughly to half the effective radius. The quality of the data and reliability
of the error bars is apparent by the typical agreement between symmetrical
points at both galaxy sides. An exception is the rotation curve for
NGC~4489. In this case, a displacement of 1.3 arcsec in the kinematic
centre can explain the asymmetry.

%%%%%%%%%%%%%%%%%%%%%%%%%%%%%%%%%%%%%%%%%%%%%%%%%%%%%%%%%%%%%%%%%%%%%%%%%%%%%%%
\section{Discussion}

\begin{figure}
\centering
\includegraphics[angle=-90,width=\columnwidth]{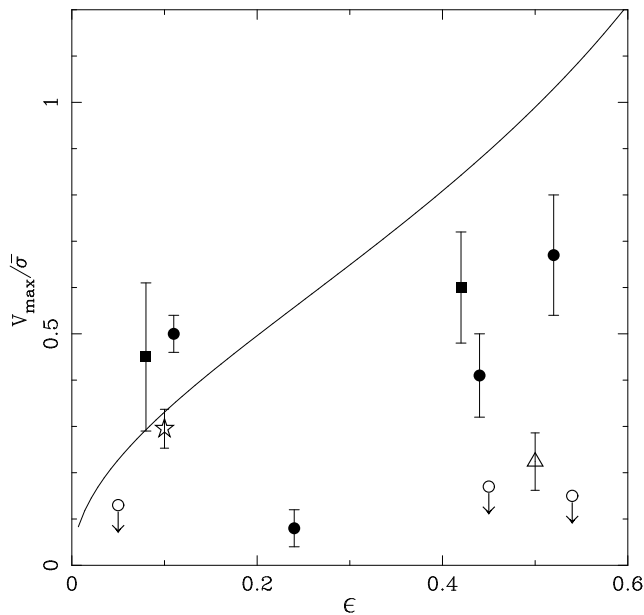}
\caption{Ratio of maximum rotational velocity to
mean velocity dispersion versus galaxy ellipticity. Filled points are the data
from this paper (circles: dEs; squares: dS0s). Open circles, the triangle and
the star show respectively data from GGV01, BN90 and DR01. The line represents
the prediction for isotropic oblate galaxies flattened by rotation from Binney
(1978).}
\label{vse}
\end{figure}

\begin{figure}
\centering
\includegraphics[angle=-90,width=\columnwidth]{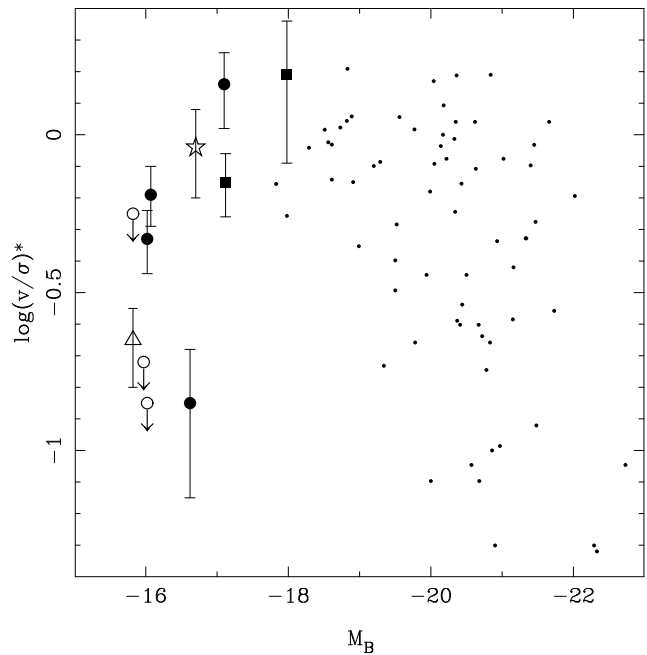}
\caption{Logarithm of the anisotropy parameter (ratio of the observed value of
$V_{\rm max}/\overline{\sigma}$ to the one predicted by the isotropic model)
against absolute magnitude. Small points show the data from Bender, Burstein
\& Faber (1992) for giant and intermediate ellipticals corrected to
$H_0=75$~km~s$^{-1}$~Mpc$^{-1}$. The rest of the symbols are the same as in
Figure~\ref{vse}.}
\label{lvsm}
\end{figure}

In order to quantify the rotational support of the objects we follow the usual
procedure of comparing the maximum rotational velocity ($V_{\rm max}$) to the
mean velocity dispersion ($\overline{\sigma}$). A conservative estimate of the
parameter $V_{\rm max}$ was computed as the error-weighted mean of the two
data pairs with the highest rotational velocities. In some galaxies, like
NGC~4431 and UGC~7436, the rotation curves increase to the outermost points
observed, so the derived $V_{\rm max}$ should be considered as lower limits to
the actual values. To compute the mean velocity dispersion we coadded all the
individual spectra with radii between $1^{\prime\prime}$ and the effective
radius. Prior to this, we shifted all the spectra to the same wavelength scale
by using the rotation curves displayed in Fig~\ref{panel}. The resulting
spectra exhibit S/N ratios, per \AA, ranging from 62 to 119.  Final values of
$V_{\rm max}$ and $\overline{\sigma}$, together with their corresponding
errors, are listed in Table~\ref{table_sample}.

In Figure~\ref{vse} we have plotted the ratio $V_{\rm max}/\overline{\sigma}$
against mean ellipticity (compiled from the most recent available estimates in
the literature -- see caption to Table~\ref{table_sample}). The previous
measurements from BN90 (for VCC~351), GGV01 (for VCC~917, VCC~1254 and
VCC~1876) and DR01 (for FS~76) are also included. For IC~794, also in the
samples of BN90 and GGV01, we only plot the ratio computed in this
work. However, it is important to note that our estimate for this galaxy
($V_{\rm max}/\overline{\sigma}=0.08\pm0.04$) is in perfect agreement with the
upper limits given by the previous works ($<0.20$ in BN90, and $<0.12$ in
GGV01). This gives further support to the data presented in this paper. We must
also note that, to be fully consistent, we have recalculated the $V_{\rm
max}/\overline{\sigma}$ ratio for the galaxy in DR01 using the same procedure
than for our sample (DR01 give a ratio of peak velocity to central velocity
dispersion of $0.33\pm0.15$). After digitizing the measurements in their
Fig.~2, we obtain a ratio of $0.30\pm0.04$. 

The curve in Figure~\ref{vse} shows the expected relationship for an oblate
spheroid with isotropic velocity distribution, thus rotationally flattened,
from Binney (1978). As it is apparent from this figure, while the dwarf
galaxies included in the previous works (with the exception of DR01) were
found to rotate too slowly to be consistent with the isotropic model, 5 out of
the 6 dwarf early-type galaxies studied in this paper (and 3 out of the 4 dEs)
are compatible with being rotationally supported. We recall that the
rotational velocities for the two galaxies at $\varepsilon\sim0.4$ could be in
fact lower limits of their actual values.

To investigate this issue in more detail, in Figure~\ref{lvsm} we present a
plot of the anisotropy parameter (ratio of the observed value of $V_{\rm
max}/\overline{\sigma}$ to the one predicted by the isotropic model) against
absolute magnitude. In order to compare the properties of dEs with those of
giant and low-luminosity Es, we have included the galaxies from the
compilation of Bender, Burstein \& Faber (1992). Concerning the three galaxies
in common with their sample (NGC~4431, IC~3393 and UGC~7436; see above), we
must note that their $V_{\rm max}/\overline{\sigma}$ values (1.00, 0.50 and
0.44, respectively) are not incompatible with ours ($0.71\pm0.16$,
$0.65\pm0.14$, $0.47\pm0.11$).

In their study of the fundamental plane, Bender et al. (1992) (see also BN90)
noted that galaxies with anisotropic velocity distributions were found at both
high and low values of mass. Apart from confirming these findings, the new
results presented in Figure~\ref{lvsm} show that the brightest early-type
dwarfs of the sample tend to be rotationally supported, extending the
behaviour of the low-luminosity ellipticals towards the bright end of the
dwarf family. Furthermore, these data indicate that, for fainter luminosities
($M_B\geq-17$), and as it was the case for the bright ellipticals, dwarf
galaxies exhibit a range of rotational properties (i.e. not all dEs are
supported by anisotropy).

In spite of these results, the present sample is still too small to derive firm
conclusions about the possible relation between rotational support and other
parameters. For instance, there are nucleated and non-nucleated dwarfs among
both the rotating and non-rotating subsamples. Note that the search for
correlations between rotational support and other properties is essential to
test the possible existence of a dichotomy within the dE family (as suggested
by Ryden et al. 1999) similar to that found for the Es (in isophote shapes,
rotation, and core properties).

It is also specially interesting to compare the rotational properties with the
stellar population of dEs. We must highlight that, with the exception of
NGC~4489 (a possible transition object between dS0 and S0), the only galaxy of
our sample flattened by anisotropic velocity dispersion (IC~794) is clearly
the most metal-rich and youngest object in G97 (see their Figure 1.b). If
confirmed by additional stellar populations measurements of other non-rotating
dEs, this still speculative result would give support to a scenario in 
which recent star
formation episodes are linked to a decrease in rotational support. In this
sense, this would agree with the suggestion by BN90 that supernova-driven
winds, triggered by star formation, could cause anisotropic supported
objects. Also, both low rotation and young stellar populations are
simultaneously predicted by the harassment models of Moore, Lake and Katz
(1998), in which cluster dE are the result of morphological transformation of
accreted spiral galaxies.  This tentative correlation between rotation and age
is in the opposite sense to that found for classical ellipticals, in which
fast-rotating Es tend to have disky isophotes and younger ages (Bender 1988;
de Jong \& Davies 1997).

In any case, and independently of the possible link between rotation and star
formation, the present results are compatible with the diversity of rotational
properties predicted by the harassment scenario (compare our Fig.~\ref{lvsm}
with Fig.~7 of Moore et al. 1998). More data is clearly needed to understand
the driving parameters governing the dynamical properties of dEs, as well as
the possible correlation of rotation with luminosity profiles, isophotal
shapes, core properties (nucleated versus non-nucleated) and stellar
populations.

%%%%%%%%%%%%%%%%%%%%%%%%%%%%%%%%%%%%%%%%%%%%%%%%%%%%%%%%%%%%%%%%%%%%%%%%%%%%%%%
\section*{Acknowledgments}

The INT is operated on the island of La Palma by the Royal Greenwich
Observatory at the Observatorio del Roque de los Muchachos of the Instituto de
Astrof\'{\i}sica de Canarias. This work was supported in part by the Spanish
Programa Nacional de Astronom\'{\i}a y Astrof\'{\i}sica under grant
AYA2000-977. 

{}

\bsp

\label{lastpage}

\end{document}